\documentclass[superscriptaddress,prl,twocolumn,nofootinbib,showpacs,preprintnumbers]{revtex4}

\usepackage{graphicx,epsfig}

\begin{document}

\title{Gauge Coupling Unification in the Standard Model}

\author{V. Barger}
\affiliation{Department of Physics, University of Wisconsin, 
Madison, WI 53706, USA}

\author{Jing Jiang}
\affiliation{Institute of Theoretical Science, University of Oregon, 
Eugene, OR 97403, USA}

\author{Paul Langacker}
\affiliation{Department of Physics and Astronomy,
University of Pennsylvania, Philadelphia, PA 19104-6396, USA}

\author{Tianjun Li}
\affiliation{School of Natural Sciences, Institute for Advanced Study,
  Einstein Drive, Princeton, NJ 08540, USA}



\begin{abstract}

The string landscape suggests that the supersymmetry breaking scale
can be high, and then the simplest low energy effective theory is 
the Standard Model (SM).
We show that gauge coupling unification can be
achieved at about $10^{16-17}$ GeV in the SM 
with suitable normalizations of the $U(1)_Y$.
Assuming grand unification scale supersymmetry breaking, 
we predict that the Higgs mass range is
127 GeV to 165 GeV, with the precise value strongly correlated with
the top quark mass and $SU(3)_C$ gauge coupling. We also present 7-dimensional 
orbifold grand unified theories in which such normalizations for the
$U(1)_Y$ and charge quantization can be realized.

\end{abstract}

\pacs{11.25.Mj, 12.10.Kt, 12.10.-g}

\preprint{MADPH-05-1421, OITS-764, UPR-1112-T, hep-ph/0503226}

\maketitle

{\bf Introduction --}  
There exists an enormous ``landscape'' for long-lived
metastable string/M theory vacua~\cite{String}. Applying
the ``weak anthropic principle''~\cite{Weinberg}, the
landscape proposal may 
be the first concrete explanation of the very tiny value of the cosmological
constant, which can take only discrete values, and it may address the gauge 
hierarchy problem.  Notably, the supersymmetry breaking scale 
can be high if there exist many supersymmetry breaking parameters or many hidden
sectors~\cite{HSUSY,NASD}.  Although there is no definite conclusion that
the string landscape predicts high-scale or TeV-scale supersymmetry 
breaking~\cite{HSUSY},
it is interesting to consider models with high-scale supersymmetry
breaking~\cite{NASD,Barger:2004sf}.

If the supersymmetry breaking scale is around the grand unification
scale or the string scale, the minimal model at low energy 
is the Standard Model (SM). The SM explains the existing experimental data very well,
 including electroweak 
precision tests, and it is easy to incorporate aspects of 
physics beyond the SM through small 
variations~\cite{NASD,Barger:2004sf,Davoudiasl:2004be}. 
However, even if the gauge hierarchy problem can be solved by the string
landscape proposal, there are still some limitations of the SM,
for example, the lack of explanation of gauge coupling unification and charge
quantization. 

Charge quantization can be easily explained by
embedding the SM into a grand unified theory (GUT). 
Should the Higgs particle be the only
new physics observed at the Large Hadron
Collider (LHC) and the SM is thus confirmed as a low energy effective theory, 
an important 
 question will be: {\it can we achieve gauge coupling 
unification in the SM without introducing any extra multiplets 
between the weak and GUT scales \cite{Frampton:1983sh}
or having large threshold corrections \cite{Calmet:2004ck}?}
As is well known, 
gauge coupling unification cannot be achieved in the SM if we
choose the canonical normalization for the $U(1)_Y$ hypercharge
interaction, {\it i.e.}, the Georgi-Glashow $SU(5)$ 
normalization \cite{Langacker:1991an}.
Also, to avoid proton decay induced by
 dimension-6 operators via heavy gauge boson exchanges,
 the gauge coupling unification scale is constrained to 
be higher than about $5\times 10^{15}$ GeV.

In this Letter we reconsider gauge coupling unification in the SM.
The gauge couplings for $SU(3)_C$ and $SU(2)_L$ are unified at
about $10^{16-17}$ GeV, and the gauge coupling for the $U(1)_Y$
at that scale depends on its normalization.
If we choose a suitable normalization
of the $U(1)_Y$, the gauge couplings for the $SU(3)_C$, $SU(2)_L$
and $U(1)_Y$ can in fact be unified at about $10^{16-17}$ GeV,
and there is no proton decay problem via dimension-6 operators.
Therefore, the real question is:
{\it  is the canonical normalization for $U(1)_Y$ unique?}

For a 4-dimensional (4D) GUT with a simple group, 
 the canonical normalization is the only possibility, assuming that
 the SM fermions form complete multiplets
under the GUT group. However, the $U(1)_Y$
normalization need not be canonical in string model 
building \cite{Dienes:1996du,Blumenhagen:2005mu},
 orbifold GUTs \cite{Orbifold,Li:2001tx}
and their deconstruction \cite{Arkani-Hamed:2001ca}, and in 4D GUTs
with product gauge groups:

(1) In weakly coupled heterotic string theory, the gauge and
gravitational couplings unify at tree
level to form one dimensionless string coupling 
constant $g_{\rm string}$ \cite{Dienes:1996du}
\begin{eqnarray}
k_Y g_Y^2 = k_2 g_2^2 = k_3 g_3^2 = 8 \pi {{G_N}\over {\alpha'}}
= g_{\rm string}^2 ~,
\end{eqnarray}
where $g_Y$, $g_2$, and $g_3$ are the gauge couplings for
the $U(1)_Y$, $SU(2)_L$, and $SU(3)_C$, respectively,
$G_N$ is the gravitational coupling
and $\alpha'$ is the string tension.
Here, $k_Y$, $k_2$ and $k_3$ are the levels of the corresponding
Kac-Moody algebras; $k_2$ and $k_3$ are positive integers while
$k_Y$ is a rational number in general \cite{Dienes:1996du}.

(2) In intersecting D-brane model building on Type II orientifolds,
the normalization for the $U(1)_Y$ (and other gauge factors) 
is not canonical in general \cite{Blumenhagen:2005mu}.

(3) In orbifold GUTs \cite{Orbifold}, the SM fermions need not form 
complete multiplets under the GUT group, so the $U(1)_Y$
normalization need not be canonical \cite{Li:2001tx}.
This statement is also valid for the deconstruction of the
orbifold GUTs \cite{Arkani-Hamed:2001ca} and for
 4D GUTs with product gauge groups.

We shall assume that at the GUT or string
scale, the gauge couplings in the SM satisfy
\begin{eqnarray}
g_1 = g_2= g_3 ~,
\end{eqnarray}
where $g_1^2 \equiv k_Y g_Y^2$, and  $k_Y=5/3$ for  
canonical normalization.
We show that gauge coupling unification in the SM can be
achieved at about $10^{16-17}$ GeV for $k_Y=4/3$, 5/4, 32/25.
Especially for $k_Y=4/3$, gauge coupling unification in the SM
is well satisfied at two loop order. Assuming GUT scale supersymmetry 
breaking, we predict that the Higgs mass is in the range 127 GeV to 165 GeV.
In addition, we construct 7-dimensional (7D) orbifold
GUTs in which such normalizations for the
$U(1)_Y$ and charge quantization can be realized. 
A more detailed discussion will be presented elsewhere \cite{BJLL}.

{\bf Gauge Coupling Unification --}
We define $\alpha_i=g_i^2/(4\pi)$ and denote the $Z$ boson
mass by $M_Z$.
In the following, we choose a top quark pole mass 
$m_t = 178.0\pm 4.3 $ GeV~\cite{Azzi:2004rc},
$\alpha_3(M_Z) = 0.1182 \pm 0.0027$~\cite{Bethke:2004uy}, and the
other gauge couplings, Yukawa couplings and the Higgs vacuum expectation
value at $M_Z$ from Ref.~\cite{Eidelman:2004wy}.

We first examine the one-loop running of the gauge couplings.
The one-loop renormalization group equations (RGEs) in the SM are
\begin{eqnarray}
(4\pi)^2\frac{d}{dt}~ g_i &=& b_i g_i^3~,~\,
\label{SMgauge}
\end{eqnarray}
where $t=\ln  \mu$, $ \mu$ is the renormalization scale, and
\begin{eqnarray}
b\equiv (b_1, b_2, b_3)=\left(\frac{41}{6 k_Y},-\frac{19}{6},-7\right)~.~\,
\label{SMbi}
\end{eqnarray}
We consider the SM with $k_Y=4/3$, 5/4, 32/25 and 5/3. In addition, we consider
the extension of the SM with two Higgs doublets (2HD) with $b=(7/k_Y, -3, -7)$ 
and $k_Y=4/3$, and the Minimal Supersymmetric
Standard Model (MSSM) with $b=(11/k_Y, 1, -3)$ and $k_Y=5/3$.
For the MSSM, we assume the supersymmetry breaking scale 300 GeV
for scenario I (MSSM I), and the effective 
supersymmetry breaking scale 50 GeV to include the threshold corrections
due to the mass differences between the squarks and sleptons for
scenario II (MSSM II) \cite{Langacker:1992rq}.
 We use $M_{U}$ to denote the unification scale where $\alpha_2$
and $\alpha_3$ intersect in the RGE evolutions.  There is a sizable
error associated with the $\alpha_3(M_Z)$ measurement.  To take into account this
uncertainty, we also consider $\alpha_3 - \delta \alpha_3$ and $\alpha_3 +
\delta \alpha_3$ as the initial values for the RGE evolutions, whose corresponding
unification scales are called $M_{U-}$ and $M_{U+}$, respectively.  
The simple relative
differences for the gauge couplings at the unification scale
are defined as $\Delta = |\alpha_1(M_{U}) - \alpha_2(M_{U})|/\alpha_2(M_{U})$, 
and $\Delta_{\pm} = |\alpha_1(M_{U\pm}) - \alpha_2(M_{U\pm})|/\alpha_2(M_{U\pm})$.

In Table~\ref{tbl:a3} we compare the convergences of the gauge couplings in above
scenarios. We confirm that the SM with canonical normalization $k_Y=5/3$
 is far from a good unification.  Introducing supersymmetry
 significantly improves the convergence.
  Meanwhile, the same level of convergences can be achieved
in all the non-supersymmetric models. 
In particular, the SM with $k_Y=32/25$ and the 2HD SM with $k_Y=4/3$
have very good gauge coupling unification.

\begin{table}[htb]
\begin{center}
\begin{tabular}{|c|c|ccc|ccc|}
\hline
Model &$k_Y$ & $M_{U-}$ & $M_{U}$ & $M_{U+}$ & $\Delta_-$ & $\Delta$ & $\Delta_+$ \\
\hline
SM & 4/3   & 1.9 & 1.4 & 1.0 & 4.3 & 3.5 & 2.6 \\
SM &  5/4   &     &     &     & 2.1 & 3.0 & 3.9 \\
SM & 32/25 &     &     &     & 0.32 & 0.60 & 1.5 \\
SM & 5/3   &     &     &     & 23.4 & 22.8 & 22.1 \\
2HD SM &  4/3 & 0.45 & 0.33 & 0.24 & 0.25 & 1.1 & 2.0 \\
MSSM I & 5/3 & 0.47 & 0.35 & 0.26 & 3.4 & 2.3 & 1.2 \\
MSSM II & 5/3 & 0.44 & 0.32 & 0.24 & 1.3 & 0.17 & 1.0 \\
\hline
\end{tabular}
\end{center}
\caption{Convergences of the gauge couplings at one loop.
  The  $M_{U}$ scales are in 
units of $10^{17}$ GeV, and the relative difference $\Delta$'s are
percentile.  }
\label{tbl:a3}
\end{table}

\begin{figure}[htb]
\centering
\includegraphics[width=8cm]{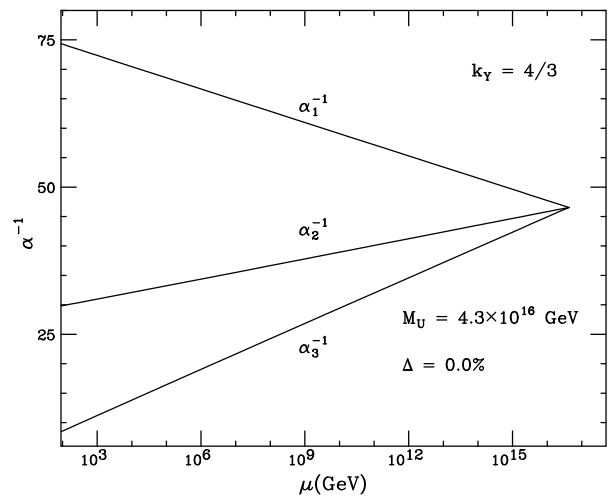}
\caption{Two-loop gauge coupling unification for the SM with $k_Y=4/3$.} 
\label{fig:2loop4o3}
\end{figure}

The two-loop running of the gauge couplings produces slightly different
results.  We perform the two-loop running for the
SM with $k_Y=4/3$, as it has an excellent unification.
We use the two-loop RGE running for the gauge
couplings and one-loop for the Yukawa couplings \cite{mac}.
With the central value of $\alpha_3$, we show the gauge coupling 
unification in Fig.~\ref{fig:2loop4o3}.
At the unification scale of $4.3 \times 10^{16}$
GeV, the value of $\alpha_1$ precisely agrees with those of $\alpha_2$
and $\alpha_3$.

\begin{figure}[htb]
\centering
\includegraphics[width=8cm]{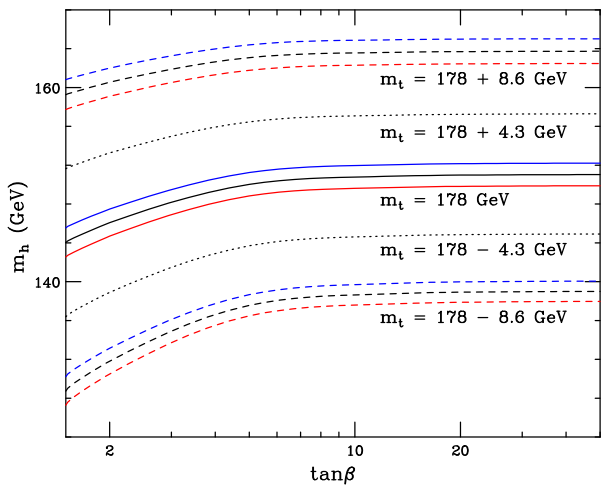}
\caption{The predicted Higgs
mass for the SM with $k_Y=4/3$. The red (lower) curves are for
$\alpha_3 + \delta\alpha_3$, the blue (upper) $\alpha_3 - \delta\alpha_3$, and
the black $\alpha_3$.  The dotted curves are for $m_t \pm \delta m_t$,
the dash ones for $m_t \pm 2 \delta m_t$, and the solid ones for $m_t$.}
\label{fig:2loopm}
\end{figure}

If the Higgs particle is the only new physics discovered at the LHC and the
SM is thus confirmed as a low energy effective theory, 
the most interesting parameter is the Higgs mass.
To be consistent with string theory or quantum gravity, it is natural
to have supersymmetry in the fundamental theory. 
In the supersymmetric models, there generically exist one pair
of the Higgs doublets $H_u$ and $H_d$.  We define the SM Higgs
doublet $H$,  which is fine-tuned to have a small mass term, 
as $H \equiv -\cos\beta i \sigma_2 H_d^*+\sin\beta H_u$, 
where $\sigma_2$ is the second
Pauli matrix and $\tan\beta $ is a mixing parameter~\cite{NASD,Barger:2004sf}.
 For simplicity, we assume that supersymmetry is broken at the GUT scale,
{\it i.e.}, the gauginos,
squarks, sleptons, Higgsinos, and the other
 combination of the scalar Higgs doublets
($\sin\beta i \sigma_2 H_d^*+\cos\beta H_u $)
have a universal supersymmetry breaking soft mass around the GUT scale.
We can calculate the Higgs boson quartic coupling $\lambda$ at the GUT
scale \cite{NASD,Barger:2004sf}
\begin{equation}
\lambda({M_U}) = \frac{k_Y g_2^2(M_U) + g_1^2(M_U)}{4
k_Y} \cos^2 2\beta~,
\end{equation}
and then evolve it down to the weak scale. 
Using the one-loop effective Higgs potential with  top quark 
radiative corrections, we calculate the Higgs boson mass by minimizing 
the effective potential \cite{Barger:2004sf}.  For the SM with $k_Y=4/3$, the
Higgs boson mass as a function of $\tan\beta$ for different $m_t$ and
$\alpha_3$ is shown in Fig.~\ref{fig:2loopm}.  We see if we vary
$\alpha_3$ within its $1\sigma$ range, $m_t$ within its $1\sigma$ and $2\sigma$
 ranges and $\tan\beta$ from
$1.5$ to $50$, the predicted mass of the Higgs boson ranges from $127$
GeV to $165$ GeV.  A large part of this uncertainty is due to the present
uncertainty in the top quark mass.  The top quark mass can be measured
to about $1$ GeV accuracy at the LHC~\cite{Beneke:2000hk}.  
Assuming this accuracy and the
central value of $178$ GeV, the Higgs boson mass is predicted to be
between $141$ GeV and $154$ GeV.

Furthermore, for the SM with $k_Y=$ 5/4 and 32/25, the gauge coupling
unifications at two loop are quite similar to that of the SM with $k_Y=4/3$,
and the predicted Higgs mass ranges are almost the same \cite{BJLL}.

{\bf Orbifold GUTs --}
In string model building, the orbifold GUTs and their deconstruction,
 and 4D GUTs with product gauge groups,
the normalization for the $U(1)_Y$ need not be canonical. As an explicit example,
we show that $k_Y=4/3$ can be obtained in the 7D orbifold $SU(6)$ model on the
space-time $M^4\times T^2/Z_6 \times S^1/Z_2$ where
 charge quantization can be realized simultaneously.
Here, $M^4$ is the 4D Minkowski space-time.
Similarly, $k_Y=5/4$ and $k_Y=32/25$ can be obtained in the 7D 
orbifold $SU(7)$ models with charge quantization \cite{BJLL}.

We consider the 7D space-time $M^4\times T^2 \times S^1$
with coordinates $x^{\mu}$, $z$ and $y$ where $z$ is the complex
coordinate for the torus $T^2$ and $y$ is the coordinate for the circle
$S^1$. The radii for $T^2$ and $S^1$ are $R$ and $R'$.
The $T^2/Z_6 \times S^1/Z_2$ orbifold is obtained from $T^2 \times S^1$ 
by moduloing the equivalent classes
\begin{eqnarray}
\Gamma_T:~~z \sim \omega  z~;~~~~~\Gamma_S:~~ y \sim -y~,~ \,
\end{eqnarray}
where $\omega =e^{{\rm i}\pi/3} $. $(z, y)=(0, 0) $ and $(0, \pi R')$ are 
the fixed points under the $Z_6\times Z_2$ symmetry.

${\cal N}=1$ supersymmetry in 7 dimensions has 16 supercharges and
 corresponds to ${\cal N}=4$ supersymmetry in 4 dimensions;
thus, only the gauge multiplet can be introduced in the bulk.  This
multiplet can be decomposed under the 4D
 ${\cal N}=1$ supersymmetry into a vector
multiplet $V$ and three chiral multiplets $\Sigma_1$, $\Sigma_2$, and 
$\Sigma_3$ in the adjoint representation, where the fifth and sixth 
components of the gauge
field ($A_5$ and $A_6$) are contained in the lowest component of $\Sigma_1$,
and the seventh component of the gauge
field ($A_7$) is contained in the lowest component of $\Sigma_2$.
The SM quarks, leptons and Higgs fields
 can be localized on 3-branes at the $Z_6\times Z_2$ fixed points,
 or on 4-branes at the $Z_6$ fixed points.

Under the $Z_6\times Z_2$ discrete symmetry, the bulk vector multiplet 
$V$ and the $\Sigma_i$ transform as \cite{BJLL}
\begin{eqnarray}
 \Phi(x^{\mu}, ~\omega z, ~\omega^{-1} {\bar z},~y) &=&
\eta_{\Phi}^T R_{\Gamma_T}
 \Phi(x^{\mu}, ~z, ~{\bar z},~y) R_{\Gamma_T}^{-1},~~\\ 
  \Phi(x^{\mu}, ~z, ~ {\bar z},~-y) &=& \eta_{\Phi}^S R_{\Gamma_S}
 \Phi (x^{\mu}, ~z, ~{\bar z},~y) R_{\Gamma_S}^{-1},~~\,
\label{SVtrans}
\end{eqnarray}
where $\Phi$ can be $V$ or $\Sigma_i$, and
\begin{eqnarray}
&& \eta_{V}^T= \eta_{\Sigma_2}^T=1,~ \eta_{\Sigma_1}^T=\omega^{-1},~
\eta_{\Sigma_3}^T=\omega~,~\\ &&
\eta_{V}^S= \eta_{\Sigma_1}^S=1,~\eta_{\Sigma_2}^S=\eta_{\Sigma_3}^S=-1~.~\,
\label{Eta}
\end{eqnarray}
We also introduce non-trivial
$R_{\Gamma_T}$ and $R_{\Gamma_S}$ to break the bulk gauge group.

Let us consider the $SU(6)$ model, which has $k_Y=4/3$.
We define the generator for the $U(1)_Y$ in $SU(6)$ as
\begin{eqnarray}
T_{U(1)_{Y}} &\equiv&  
{\rm diag}\left( {1\over 3}, {1\over 3}, {1\over 3}, 
- {1\over 3}, - {1\over 3}, - {1\over 3} \right)~.~\,
\label{GU1Y}
\end{eqnarray}
Because ${\rm tr} [T_{U(1)_{Y}}^2]=2/3$, we obtain $k_Y=4/3$.

To break the $SU(6)$ gauge symmetry, we choose
the following $6\times 6$ matrix representations for 
$R_{\Gamma_T}$ and $R_{\Gamma_S}$
\begin{eqnarray}
R_{\Gamma_T} &=& {\rm diag} \left(+1, +1, +1,
 \omega^{2}, \omega^{2}, \omega^{5} \right)~,~ \\
R_{\Gamma_S} &=& {\rm diag} \left(+1, +1, +1, +1, +1, -1 \right)~.~\,
\end{eqnarray}
We obtain that, for the zero modes, the 7D 
${\cal N} = 1 $ supersymmetric $SU(6)$ gauge symmetry is broken 
down to the 4D ${\cal N}=1$ supersymmetric
$SU(3)_C\times SU(2)_L\times U(1)_Y \times U(1)'$ \cite{BJLL}.
Also, we have only one zero mode from $\Sigma_i$ with
quantum number $\mathbf{(\bar 3, 1, -2/3)}$ under the SM gauge symmetry,
which can be considered as the right-handed top quark \cite{BJLL}.

On the 3-brane at the $Z_6\times Z_2$ fixed point
 $(z, y)= (0, 0)$, the preserved gauge symmetry 
is $SU(3)_C\times SU(2)_L\times U(1)_Y \times U(1)'$ \cite{Li:2001tx}.
Thus, on the observable 3-brane at $(z, y)= (0, 0)$, 
we can introduce  one pair of Higgs doublets and
three families of SM quarks and leptons except
the right-handed top quark \cite{Li:2001tx}. Because the $U(1)_Y$ charge 
for the right-handed top quark is determined from the construction,
charge quantization can be achieved from the anomaly free
conditions and the gauge invariance
of the Yukawa couplings on the observable 3-brane.
Moreover, the $U(1)'$ anomalies can be cancelled by
assigning suitable $U(1)'$ charges to the SM 
quarks and leptons, and
the $U(1)'$ gauge symmetry can be broken at the GUT
scale by introducing
one pair of the SM singlets with $U(1)'$ charge $\pm1$ 
on the observable 3-brane.
Interestingly, this $U(1)'$ gauge symmetry may be considered
as a flavour symmetry, and then the fermion masses and
mixings may be explained naturally.
Furthermore, supersymmetry can be broken around the 
compactification scale, which can be considered as the GUT scale, 
for example, by Scherk--Schwarz mechanism \cite{Scherk:1978ta}.

We briefly comment on the 7D orbifold $SU(7)$ models which 
can have $k_Y=5/4$ and $k_Y=32/25$ \cite{BJLL}.
The discussion is similar to that for the above $SU(6)$ model.
The 7D ${\cal N} = 1 $ supersymmetric $SU(7)$ gauge symmetry is broken 
down to the 4D ${\cal N}=1$ supersymmetric
$SU(3)_C\times SU(2)_L\times U(1)_Y \times U(1)'\times U(1)''$
 by orbifold projections. There is only
 one pair of  zero modes from $\Sigma_i$ with
quantum numbers $\mathbf{(1, 2, +1/2)}$
and $\mathbf{(1, 2, -1/2)}$ under the SM gauge symmetry,
which can be considered as one pair of Higgs doublets.
Also, charge quantization can be realized \cite{BJLL}.

{\bf Conclusions --}
The string landscape suggests that the supersymmetry breaking scale
can be high and then the simplest low energy effective theory is
just the SM.
We showed that gauge coupling unification in the SM 
with $k_Y$=4/3, 5/4, and 32/25 can be
achieved at about $10^{16-17}$ GeV.
Assuming  GUT scale supersymmetry breaking,
 we predicted that the Higgs mass is
in the range 127 GeV to 165 GeV. We also presented the 7D
orbifold GUTs where such normalizations for the
$U(1)_Y$ and charge quantization can be realized.

{\bf Acknowledgments --}
 This research was supported by the U.S.~Department of Energy
under Grants No.~DE-FG02-95ER40896, DE-FG02-96ER40969 and DOE-EY-76-02-3071,
 by the National Science
Foundation under Grant No.~PHY-0070928, and by the University
of Wisconsin Research Committee with funds granted by the Wisconsin
Alumni Research Foundation.


\end{document}